\documentclass[a4paper,fleqn]{cas-dc}

\usepackage[numbers]{natbib}
\usepackage{color}

\def\tsc#1{\csdef{#1}{\textsc{\lowercase{#1}}\xspace}}
\tsc{WGM}
\tsc{QE}

\begin{document}
\let\WriteBookmarks\relax
\def\floatpagepagefraction{1}
\def\textpagefraction{.001}

\shorttitle{Control of fragment sizes of exploding rings}    

\shortauthors{C. Szuszik and F. Kun}  

\title [mode = title]{Control of fragment sizes of exploding rings}  



%

\author[1]{Csan\'ad Szuszik}[orcid=0009-0000-6353-1421]



\ead{csanad.szuszik@science.unideb.hu}



\affiliation[1]{organization={Department of Theoretical Physics, Doctoral School of Physics, Faculty of Science and Technology, University of Debrecen},
            addressline={P.O. Box 400}, 
            postcode={H-4002}, 
            postcodesep={}, 
            city={Debrecen},
            country={Hungary}}

\author[1,2]{Ferenc Kun}[orcid=0000-0001-6469-7917]

\cormark[1]


\ead{ferenc.kun@science.unideb.hu}



\affiliation[2]{organization={HUN-REN Institute for Nuclear Research, HUN-REN ATOMKI},
            addressline={P.O. Box 51}, 
            postcode={H-4001}, 
            postcodesep={}, 
            city={Debrecen},
            country={Hungary}}

\cortext[1]{Corresponding author}


\nonumnote{Supported by the \'UNKP-23-3 New National Excellence Program of the Ministry for Culture and Innovation from the source of the National Research, Development and Innovation Fund. Supported by the University of Debrecen Program for Scientific Publication. Project no. TKP2021-NKTA-34 has been implemented with the support provided from the National Research, Development and Innovation Fund of Hungary, financed under the TKP2021-NKTA funding scheme.}

\begin{abstract}
We investigate the fragmentation of ring-like brittle structures under explosive loading using a discrete element model. By systematically varying ring thickness and strain rate, we uncover a transition from one-dimensional (1D) segmentation to two-dimensional (2D) planar fragmentation and, ultimately, to complete shattering. This transition is driven by the effective dimensionality of the crack pattern, which evolves with increasing strain rate. We identify a critical ring thickness beyond which segmentation ceases, and fragmentation directly follows a power-law mass distribution characteristic of 2D systems. In the crossover regime, spanning and non-spanning fragments coexist, enabling control over the power-law exponent of the mass distribution. At very high strain rates, we observe a transition to complete shattering, where the system follows a novel scaling law relating the shattering strain rate to ring thickness. Our results provide fundamental insights into fragmentation universality classes and offer potential applications in space debris prediction, controlled detonation technologies, and materials engineering.
\end{abstract}




\begin{keywords}
shell fragmentation \sep dimensional crossover \sep phase transition \sep scaling laws
\end{keywords}

\maketitle

\section{Introduction}
Energetic loading of solid bodies such as impact or explosion results in a sudden disintegration into numerous pieces with a large variety of shapes and sizes \cite{astrom_statistical_2006,bonn_natcomm_2021,turcotte_factals_1986,steacy_automaton_1991,turcotte_fractals_1997,brilliantov_size_2015}. Such dynamic fragmentation phenomena are ubiquitous in nature, from the pyroclastic activity of volcanic eruptions through the collapse of rock walls to the breakup of glaciers \cite{edwards2020,astrom_termini_2014,astrom_glaciology_2021} they often occur in our geological environment.
Beyond understanding patterns in nature \cite{turcotte_factals_1986,steacy_automaton_1991,brilliantov_size_2015,Domokos18178}, one of the most important challenges of fragmentation research is to predict and control the size (mass) distribution of the generated pieces \cite{astrom_statistical_2006,bonn_natcomm_2021}. For dynamic fragmentation, universality classes have been identified with distinct mass distributions, i.e.\ exponential and power law mass distributions have been obtained when breakup is fueled by internal stresses \cite{bonn_natcomm_2021} or externally imposed loading \cite{oddershede_self-organized_1993, astrom_universality_2000,astroem_exponential_2004,astroem_universal_2004,wittel_fragmentation_2004,kun_scaling_2006,timar_new_2010}, respectively. Experiments and computer simulations revealed a high degree of robustness of fragment mass distributions in universality classes against details of material properties and loading conditions, leaving only a little room for control \cite{astroem_exponential_2004,astroem_universal_2004,kun_transition_1999,wittel_fragmentation_2004,kun_scaling_2006,katsuragi_scaling_2003,katsuragi_crossover_2004, timar_new_2010}. Typically, global characteristics of the fragment population such as the average mass, and the upper cutoff of fragment masses can be tuned e.g.\ with the amount of imparted energy, however, functional forms and the value of power law exponents proved to be strongly resistant \cite{timar_new_2010,katsuragi_scaling_2003,PhysRevE.86.016113,clemmer_prl_2022,clemmer_pre_2023}. This robust behaviour could be cast into scaling laws which motivated the concept of the analogy of fragmentation processes to critical phenomena \cite{kun_transition_1999, katsuragi_scaling_2003, wittel_fragmentation_2004, kun_scaling_2006, astroem_exponential_2004, astroem_universal_2004, timar_new_2010}.

Here, we show that by varying the imparted energy, a transition between universality classes of fragmentation can be achieved by altering the effective dimensionality of crack patterns. Using a discrete element model of heterogeneous materials, we explore how circular rings embedded in a two-dimensional plane break under explosive forces at different thicknesses by varying the initial strain rate. At low strain rates, rings break into fragments through radial cracks, resulting in a one-dimensional breakup and a Weibull-type fragment mass distribution. However, beyond a critical strain rate, cracks propagate throughout the entire ring, leading to a structural transition to planar fragmentation with a universal power-law mass distribution. By isolating different fragment types, we reveal that this dimensional crossover allows control of the power-law exponent of mass distributions up to a critical thickness. At high strain rates, a second transition occurs, resulting in complete shattering at all thicknesses, characterized by a distinct scaling law. Our findings are crucial not only for their theoretical implications but also for predicting the size distribution of space debris, such as that generated by exploding rocket fuel containers \cite{space_debris_science_2006,space_debris_impact_2008}, and for precisely controlling fragment mass distribution in exploding shells for advanced defense technologies \cite{grady_dynamic_2010,zhang_shell_2018}.

\section{Discrete element simulation of the explosion of rings}
\begin{figure}
\centering\includegraphics[width=3.2in]{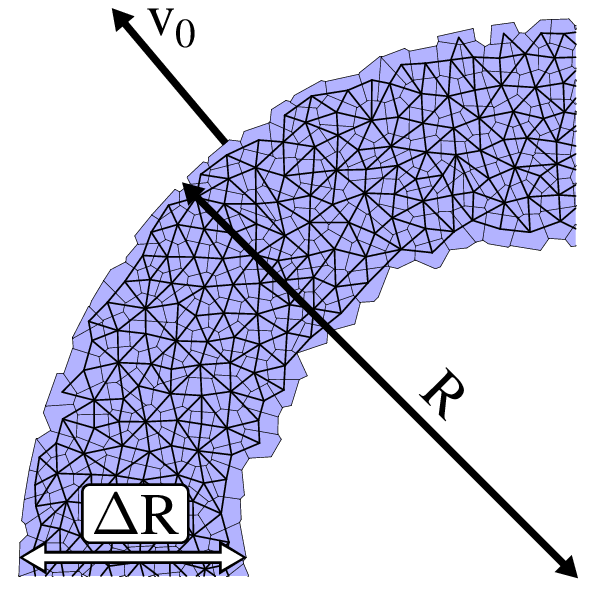}
\caption{\label{fig:modelillust} \textcolor{black}{Model setup illustrating the discretization of a quarter of a ring into convex polygons. Neighboring polygons of the tessellation are connected by beam elements (thicker black lines). The outer radius of the sample is $R=60l_0$, and its relative thickness has the value $\Delta R/R= 0.35$. The initial speed $\vec{v}_0$ of polygons points radially outward.} }
\end{figure}

To study the explosion-induced breakup of brittle rings, we employ a two-dimensional discrete element model (DEM) that captures the essential aspects of solid fragmentation, including the heterogeneous microstructure, mechanical behavior, and fracturing dynamics \cite{kun_study_1996,colliding_disc_1996, kun_transition_1999}.
In our DEM, we discretize a planar ring by generating a Voronoi tessellation of a rectangular sample and then cutting out two concentric circles with radii $R$ and $R-\Delta R$, where $\Delta R$ is the ring thickness. \textcolor{black}{The random structure of the polygonal lattice provides a high degree of isotropy of the discretization.} To ensure good spatial resolution across thicknesses, the outer radius is fixed at $R=120 l_0$, with $\Delta R$ varied to cover a ratio range of $\Delta R/R=0.02-1.0$. (Here $l_0$ denotes the characteristic polygon size of the tessellation. Thus, the smallest thickness is $\Delta R=2.5 l_0$, while at the highest thickness $\Delta R/R=1$ the system comprises approximately 50,000 Voronoi polygons. \textcolor{black}{Figure \ref{fig:modelillust} illustrates the model construction. Individual polygons represent mesoscopic material elements, which have three degrees of freedom in two dimensions, i.e.\ the two coordinates of the center of mass, and a rotation angle.}

In the initial tessellation, we connect neighboring polygons' centers of mass with beam elements that exert forces and torques when elongated, compressed, sheared, or bent \cite{kun_study_1996, colliding_disc_1996, kun_transition_1999}. \textcolor{black}{Geometrical properties of beams such as their length and cross-section derive from the random tessellation, introducing structural disorder in the solid.} During deformation beams break according to a criterion which takes into account that stretching and bending (shear) contribute to breaking \cite{kun_study_1996,colliding_disc_1996, kun_transition_1999}
\begin{align}
    \left(\frac{\varepsilon_{ij}}{\varepsilon_{th}}\right)^2+\frac{\max{\left(\left|\Theta_i\right|,\left|\Theta_j\right|\right)}}{\Theta_{th}}>1.
    \label{eq:break}
\end{align}
Here $\varepsilon_{ij}$ denotes the longitudinal strain of the beam connecting polygons $i$ and $j$, while $\Theta_i$ and $\Theta_j$ are the bending angles at the two beam ends. \textcolor{black}{The relative importance of the breaking modes is controlled by the two breaking parameters $\varepsilon_{th}$ and $\Theta_{th}$, which have fixed values for all the beams. However, beam breaking still has a certain degree of randomness, because the structural disorder introduced by the random Voronoi tessellation shows up also in the mechanical properties of beams} \cite{daddetta_application_2002,kun_study_1996}.

Explosive loading is simulated by assigning each polygon an initial radial velocity $v_0$, and the system's evolution is tracked by solving the equations of motion for the polygons' translational and rotational degrees of freedom using a 5th order Predictor-Corrector scheme \cite{allen_computer_1984}. The breaking criterion Eq.\ (\ref{eq:break}) is evaluated at each iteration, removing over-stressed beams that meet the breaking condition. At adjacent broken beams the edges of polygons form extended cracks, along which the ring fractures and falls apart.
Contact between unconnected polygons generates a repulsive force proportional to the overlap area that prevents crack faces to penetrate each other \cite{colliding_disc_1996,kun_study_1996,kun_transition_1999}.

Our model has been successfully applied to describe the dynamical and statistical features of fracture and fragmentation in heterogeneous brittle materials under various loading conditions \cite{colliding_disc_1996,kun_study_1996,kun_transition_1999}. Further details of the model construction and the parameter values used in the simulations can be found in Refs.\ \cite{addetta_solids_2001,daddetta_application_2002,colliding_disc_1996,kun_study_1996,kun_transition_1999}. Explosive loading is characterized by the strain rate $\dot{\varepsilon}$, controlled by the initial polygon speed $v_0$ as $\dot{\varepsilon}=v_0/R$. This strain rate is normalized by $v_e/R$, where $v_e$ is the speed of elastic waves in the model solid, determined numerically.

\section{Phase diagram of exploding rings}
\begin{figure}
\centering\includegraphics[width=3.3in]{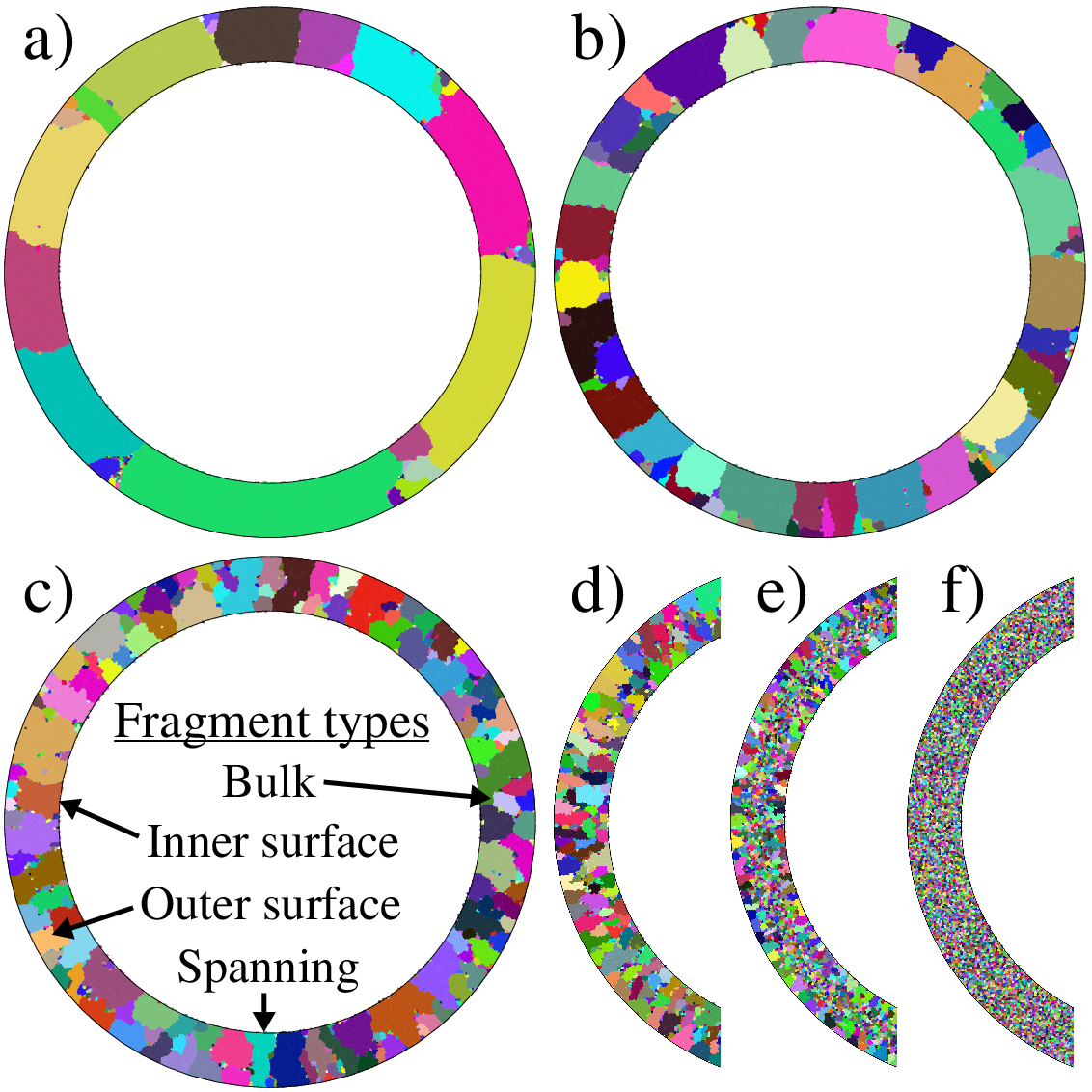}
\caption{\label{fig:finstate} Final states of fragmenting rings at the thickness $\Delta{R}/R=0.2$ for different strain rates $\dot{\varepsilon}$: $8\times10^{-5}$ $(a)$, $1\times10^{-4}$ $(b)$, $3\times10^{-4}$ $(c)$, $8\times10^{-4}$ $(d)$, $2\times10^{-3}$ $(e)$, $4\times10^{-2}$ $(f)$. Fragments are highlighted by randomly assigned colors and they are placed back to their original position inside the ring. In $(c)$ examples of different types of fragments are highlighted.}
\end{figure}

Computer simulations of the explosion of rings revealed that crack initiation requires the strain rate $\dot{\varepsilon}$ to exceed a material-dependent threshold value $\dot{\varepsilon}_c$. Figure \ref{fig:finstate} shows the final states of rings exploding at different $\dot{\varepsilon}>\dot{\varepsilon}_c$ values, with a fixed thickness of $\Delta{R}/R=0.2$, illustrating fragments in their original positions. Due to the expansion of the system, cracks typically begin at the external surface and propagate inward, segmenting the ring in an essentially one-dimensional way (Fig.\ \ref{fig:finstate}$(a)$). As $\dot{\varepsilon}$ increases, the crack spacing decreases, reducing fragment size (see Fig.\ \ref{fig:finstate}$(b)$), consistent with Mott and Grady's predictions \cite{mott_fragmentation_1947, grady_geometric_1985}, indicating that our dynamical model correctly captures the interplay of the strain rate of loading, the heterogeneity of the material, and the stress release waves. At higher $\dot{\varepsilon}$ values, the breakup scenario substantially changes: elastic wave interference creates a highly stretched region along the ring's middle, activating secondary cracks perpendicular to radial direction and producing fragments that no longer span the ring from the inner to the outer surface (Fig.\ \ref{fig:finstate}$(c)$). Further increasing $\dot{\varepsilon}$ results in a more complex crack structure due to the growing number of secondary cracks: the ring segments gradually disappear, although the elongated shape of pieces is still reminiscent of the strong radial orientation of primary cracks (Fig.\ \ref{fig:finstate}$(d)$). In the limit of very high $\dot{\varepsilon}$ values the high density of cracks results in the complete shattering of the ring into powder, i.e.\ single polygons in our DEM (Figs.\ \ref{fig:finstate}$(e,f)$).
\begin{figure}
 \centering\includegraphics[width=3.4in]{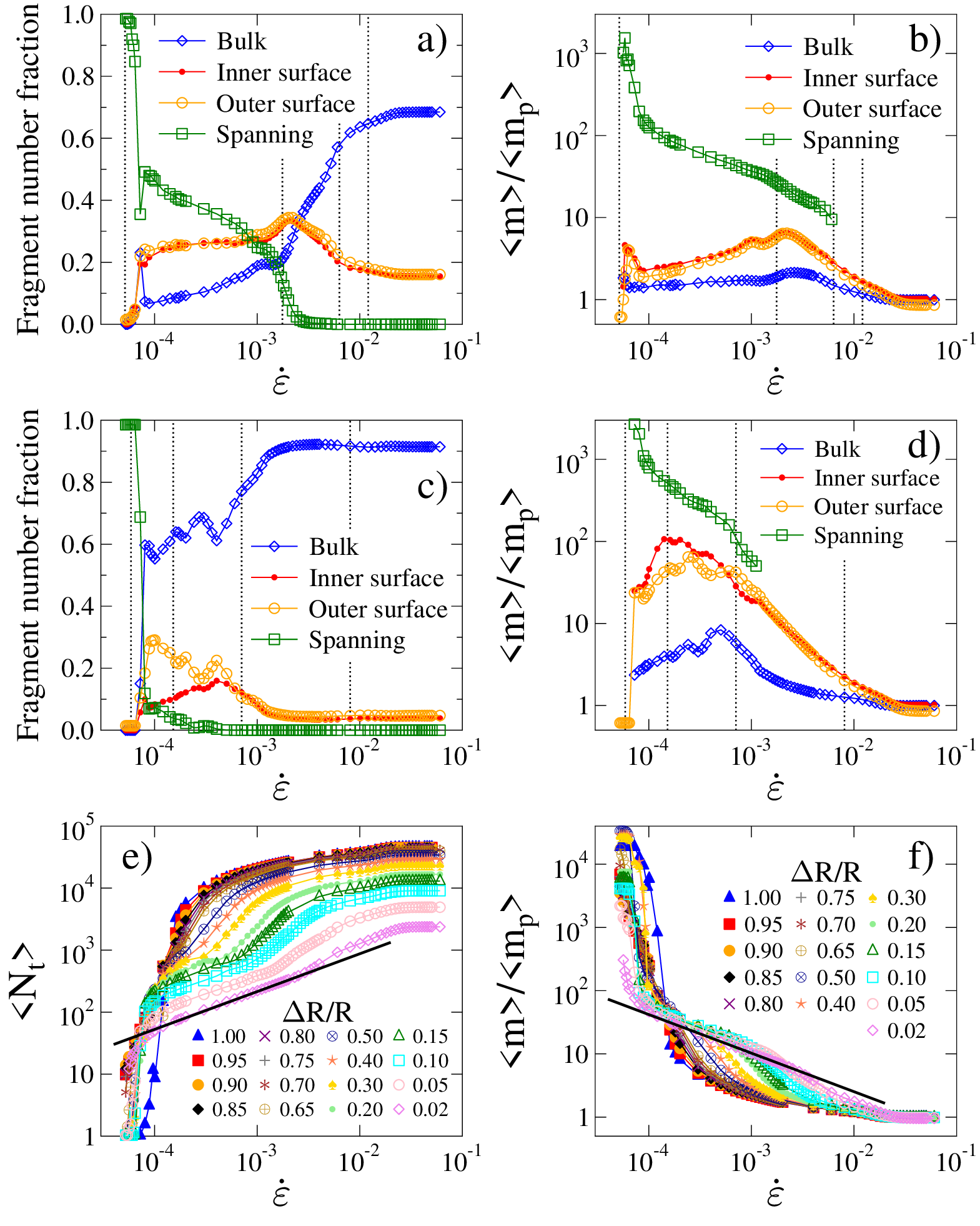}
 \caption{\label{fig:fragnummass} Average number $(a)$ and mass $(b)$ of fragments of different types as function of strain rate $\dot{\varepsilon}$ for the ring thickness $\Delta R/R=0.05$.  \textcolor{black}{$(c,d)$ present the same quantities for a higher thickness $\Delta R/R=0.2$.} Average total number of fragments $(e)$ and their average mass $(f)$ are presented as function of $\dot{\varepsilon}$ for all the thicknesses considered without separating different fragment types. \textcolor{black}{In $(a,b)$ and $(c,d)$ the vertical dotted lines indicate the corresponding values of the characteristic strain rates: $\dot{\varepsilon}_c=  5.15 \times 10^{-5}$, $\dot{\varepsilon}_{sc} = 1.76 \times 10^{-3}$, $\dot{\varepsilon}_f = 6.33 \times 10^{-3}$, and $\dot{\varepsilon}_s =  1.21 \times 10^{-2}$ for $(a,b)$, and $\dot{\varepsilon}_c=  5.89 \times 10^{-5}$, $\dot{\varepsilon}_{sc} = 1.52 \times 10^{-4}$, $\dot{\varepsilon}_f = 7.06 \times 10^{-4}$, and $\dot{\varepsilon}_s =  8.06 \times 10^{-3}$ for $(c,d)$, from left to right}. In $(e)$ and $(f)$ the straight lines represent power laws of exponent $\alpha=0.61$. $\left<m_p\right>$ denotes the average mass of single polygons. In all the figures the same range of $\dot{\varepsilon}$ is used on the horizontal axis.}
\end{figure}

To unveil how the evolution of the crack structure determines the emerging fragments, in the final state we classify them as spanning, bulk, or surface pieces. Spanning fragments extend radially, bulk fragments lie entirely within the ring, and surface fragments reach either the inner or outer surface (see Fig.\ \ref{fig:finstate}$(c)$ for examples of fragment types). Figure \ref{fig:fragnummass}$(a)$ shows for a thin ring $\Delta R/R=0.05$ the evolution of the number of each fragment type $N_s$ (spanning), $N_b$ (bulk), $N_{in}$, $N_{out}$ (inner and outer surface) normalized by the total fragment number $N_{t}$ and averaged over 150 simulations at all $\dot{\varepsilon}$ values. Just above the threshold strain rate $\dot{\varepsilon}_c$, most fragments are spanning segments formed by radial cracks $\left<N_s/N_t\right>\approx 1$ (see also Fig.\ \ref{fig:finstate}$(a)$). As $\dot{\varepsilon}$ increases, small pieces break off the spanning ones so that the proportion of spanning fragments $\left<N_s/N_t\right>$ decreases, while bulk $\left<N_b/N_t\right>$ and surface fragments $\left<N_{in}/N_t\right>$, $\left<N_{out}/N_t\right>$ increase. The average mass of different fragment types, shown in Fig.\ \ref{fig:fragnummass}$(b)$, indicates that initially bulk and surface fragments comprise only a few polygons, whereas spanning segments are orders of magnitude larger. As a consequence of the evolution of the crack structure (Fig.\ \ref{fig:finstate}), a threshold strain rate, $\dot{\varepsilon}_{sc}$, emerges which marks a significant change in fragment behavior: spanning fragments nearly vanish, while the surface fragment fraction peaks and the fraction of bulk fragments sets to a faster increase (Fig.\ \ref{fig:fragnummass}$(a)$), accompanied by a maximum in their average mass (Fig.\ \ref{fig:fragnummass}$(b)$). This behaviour suggests the onset of a transition from one-dimensional (1D) radial segmentation to planar (2D) fragmentation as $\dot{\varepsilon}$ increases due to the activation of internal degrees of freedom of the ring.
Our calculations revealed that the initiation point of this dimensional crossover can be quantitatively identified as the strain rate $\dot{\varepsilon}_{sc}$, where the majority of the fragment mass is first comprised by non-spanning (bulk and surface) fragments.
The transition to planar fragmentation is completed at the strain rate $\dot{\varepsilon}_f$, where the number of spanning fragments drops to zero $\left<N_s/N_t\right>=0$. It can be observed in Fig.\ \ref{fig:fragnummass}$(a,b)$ that at the characteristic strain rates $\dot{\varepsilon}_{c}$, $\dot{\varepsilon}_{sc}$, and $\dot{\varepsilon}_{f}$, indicated by the vertical dotted lines, both the average number and mass of different fragment types undergo qualitative changes. \textcolor{black}{To illustrate the influence of ring thickness on the fragmentation behavior, Figs.\ \ref{fig:fragnummass}$(c,d)$ show the same quantities as in Figs.\ \ref{fig:fragnummass}$(a,b)$, but for a higher relative thickness of $\Delta R/R=0.2$. While the overall trends in the fraction and average mass of different fragment types remain qualitatively similar with increasing strain rate, a notable increase is observed in the fraction of bulk fragments $\left<N_b/N_t\right>$ (Fig.\ \ref{fig:fragnummass}$(c)$). It is also worth noting that the segmentation phase becomes significantly narrower, whereas the fragmentation phase broadens at this higher ring thickness.}

\begin{figure}
\centering\includegraphics[width=3.4in]{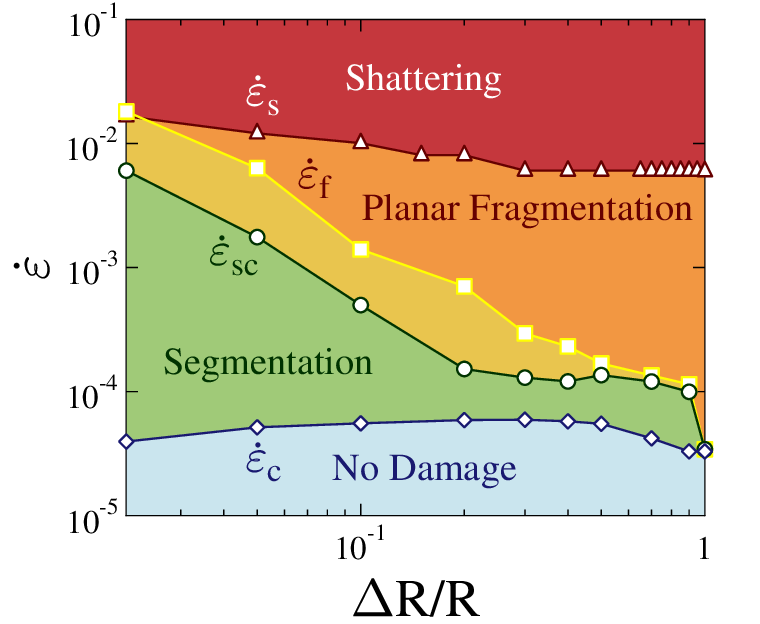}
\caption{Phase diagram of exploding rings on the thickness-strain rate plane. Cracking starts at the threshold $\dot{\varepsilon}_c$ leading to segmentation. The breakup scenario changes to planar fragmentation at $\dot{\varepsilon}_f$ after a crossover regime between $\dot{\varepsilon}_{sc}$ and $\dot{\varepsilon}_f$, where spanning segments and non-spanning fragments coexist. Complete shattering sets on at the threshold strain rate $\dot{\varepsilon}_s$.  \label{fig:phasediag}}
\end{figure}

To provide a comprehensive overview of the system's complexity, Figure \ref{fig:phasediag} presents the phase diagram of exploding rings, mapped onto the thickness–strain rate plane, $\Delta R/R-\dot{\varepsilon}$. Our calculations showed that segmentation begins at nearly the same threshold strain rate $\dot{\varepsilon}_c$ at all ring thicknesses $\Delta R/R$. However, the phase boundaries, $\dot{\varepsilon}_{sc}$ and $\dot{\varepsilon}_{f}$, decrease with increasing $\Delta R/R$ in such a way that both the segmentation phase and the crossover regime, where spanning and non-spanning fragments coexist, progressively narrow. Importantly, our findings confirm the existence of a critical thickness $\Delta R_c/R\approx 0.3$ beyond which the segmentation phase becomes very narrow, while the crossover regime entirely vanishes at $\Delta R/R\approx 0.5$.
Note that in the limit $\Delta R/R \to 1$ no segmentation can occur in the model.

\section{Evolution of the fragment mass distribution}
The statistics of fragment masses is very sensitive to the effective dimensionality of the fracturing system \cite{astrom_universality_2000,astroem_exponential_2004,wittel_fragmentation_2004}. For fragmenting rings, analytical calculations have predicted an asymptotic power law decrease of the average mass of fragments $\left<m\right>$ as
\begin{align}
\left<m\right> \sim \dot{\varepsilon}^{-\alpha},
\label{eq:avermass}
\end{align}
where the exponent $\alpha$ universally equals $2/3$ for heterogeneous brittle materials \cite{mott_fragmentation_1947,grady_geometric_1985}. For our simulations, Figure \ref{fig:fragnummass}$(f)$ shows the average fragment mass $\left<m\right>$ (without distinguishing their type) for various thicknesses $\Delta R/R$ as a function of strain rate. The power law relation in Eq.\ (\ref{eq:avermass}) holds with high accuracy over the strain rate range between $\dot{\varepsilon}_c$ and $\dot{\varepsilon}_{sc}$ of the phase diagram Fig.\ \ref{fig:phasediag}, where spanning segments dominate in the fragment mass. Note that the power law regime shrinks with increasing ring thickness in agreement with the evolution of the phase boundaries. The exponent obtained by fitting, $\alpha=0.61\pm 0.04$, closely matches the analytical prediction. Consequently, the total number of fragments, shown in Fig.\ \ref{fig:fragnummass}$(e)$, increases monotonically following a power law $\left<N_t\right>\sim \dot{\varepsilon}^{\alpha}$, in the corresponding strain rate range with the same exponent $\alpha$ as the average fragment mass, as expected.

\begin{figure}
\centering\includegraphics[width=3.4in]{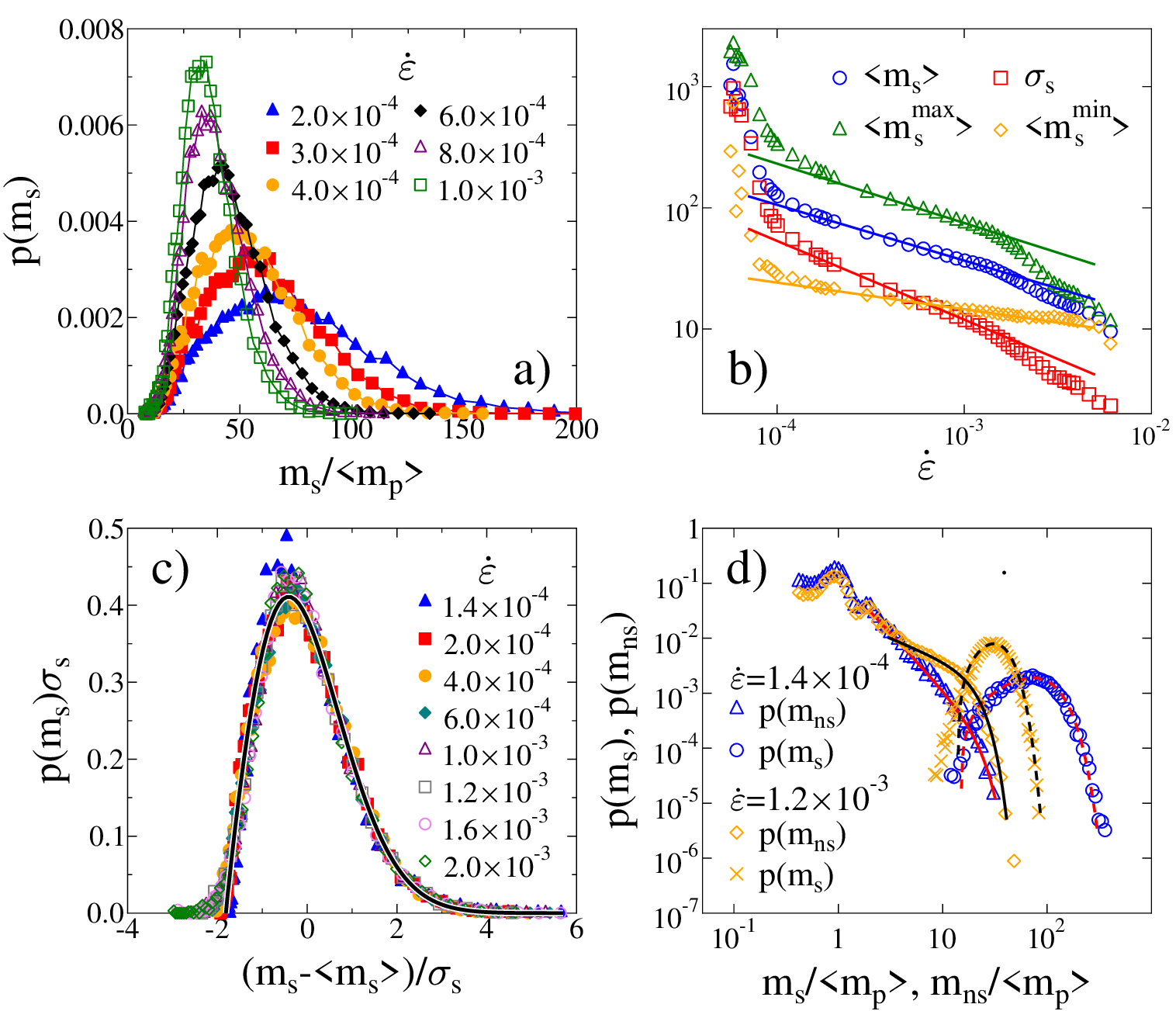}
\caption{\label{fig:lognorm} $(a)$ Mass distributions of spanning fragments $p(m_s)$ at the thickness $\Delta{R}/R=0.05$ for several strain rates in the range $\dot{\varepsilon}_c<\dot{\varepsilon}<\dot{\varepsilon}_f$. $(b)$ The average mass of spanning fragments $\left<m_s\right>$, the average mass of the smallest $\left<m_s^{min}\right>$ and largest $\left<m_s^{max}\right>$ pieces, and the standard deviation $\sigma_s$ as a function of strain rate for the same thickness as in $(a)$. $(c)$ Data collapse analysis of the mass distribution of spanning fragments.
The continuous line represents the best fit of the scaling function obtained with Eq.\ (\ref{eq:weibull}). $(d)$ Coexistence of Weibull type distribution of spanning and power law distribution of non-spanning fragments demonstrated for two strain rates at the thickness $\Delta R/R=0.05$.}
\end{figure}

In order to isolate the contributions of different cracking mechanisms to the statistics of fragments, we determined the fragment mass distribution separately for spanning and non-spanning pieces. Figure \ref{fig:lognorm}$(a)$ demonstrates for the thickness $\Delta R/R=0.05$ that until spanning fragments survive, i.e.\ between $\dot{\varepsilon}_c$ and $\dot{\varepsilon}_f$ (see Fig.\ \ref{fig:phasediag}), their mass distribution $p(m_s)$ has a very robust functional form. It can be seen in Fig.\ \ref{fig:lognorm}$(b)$ that increasing $\dot{\varepsilon}$, not only the average mass of spanning fragments $\left<m_s\right>$, but the average mass of the smallest $\left<m_s^{min}\right>$ and largest $\left<m_s^{max}\right>$ pieces, and the standard deviation $\sigma_s$ of the distribution decrease according to the power law functional form of Eq.\ (\ref{eq:avermass}), although with slightly different exponents $\alpha_{av}=0.47$, $\alpha_{min}=0.22$, $\alpha_{max}=0.49$, $\alpha_{sd}=0.65$, respectively. The results indicate that while the distributions shift to lower masses in Fig.\ \ref{fig:lognorm}$(a)$ they get narrower but in such a way that the ratio of the largest and the average remains nearly the same. Beyond the power-law regime, both characteristic fragment masses, $\left<m_s\right>$ and $\left<m_s^{max}\right>$, converge to $\left<m_s^{min}\right>$, the lower cutoff of the mass distribution $p(m_s)$. Simultaneously, the standard deviation $\sigma_s$ approaches zero, signaling the gradual disappearance of spanning fragments.

Figure \ref{fig:lognorm}$(c)$ presents that the distributions $p(m_s)$ obtained at different strain rates could be collapsed on a master curve by rescaling with the average mass $\left<m_s\right>$ and standard deviation $\sigma_s$.
The mass distributions could be very well fitted by a Weibull type distribution shifted with the minimum mass $m_s^{min}$
\begin{align}
    p(m_s)=\mu \frac{(m_s+m_s^{min})^{\mu-1}}{\lambda^{\mu}}\exp{\left[-\left(\frac{m_s+m_s^{min}}{\lambda}\right)^{\mu}\right]}.
    \label{eq:weibull}
\end{align}
Here the exponent $\mu$ controls the shape of the distribution $p(m_s)$, while $\lambda$ and $m_s^{min}$ set the scale of fragment masses. \textcolor{black}{It is important to emphasize that in Figure \ref{fig:lognorm}$(c)$ the numerically obtained  parameter values $\mu=1.93$ and $\lambda=2.04$ of the fitted curve are consistent with a Rayleight distribution predicted in Ref.\ \cite{molinari_ring_apl2006}. The functional form Eq.\ (\ref{eq:weibull}) of the mass distribution together with the power law decay of the average fragment mass Eq.\ (\ref{eq:avermass}) provide a reasonable description of the outcomes of ring fragmentation experiments \cite{grady_geometric_1985,ring_warhead_simul_2015}.} However, the results imply that the validity of the Rayleight distribution goes beyond the segmentation phase, it describes the statistics of spanning fragments up to the fragmentation threshold $\varepsilon_f$.

Our calculations revealed that for non-spanning fragments the mass distribution $p(m_{ns})$ has a power law functional form followed by an exponential cutoff
\begin{align}
    p(m_{ns})\sim m_{ns}^{-\tau}\exp{(-m_{ns}/m_0)},
    \label{eq:powlaw}
\end{align}
both below and above the critical point $\dot{\varepsilon}_f$ of planar fragmentation. The difference is that for $\dot{\varepsilon}< \dot{\varepsilon}_f$, where spanning fragments still exist, the power law exponent $\tau$ depends on the explosion strain rate $\dot{\varepsilon}$. Figure \ref{fig:lognorm}$(d)$ demonstrates the coexistence of the Weibull type mass distribution of ring segments Eq.\ (\ref{eq:weibull}) and the power law distribution Eq.\ (\ref{eq:powlaw}) of non-spanning fragments at two strain rates below $\dot{\varepsilon}_f$. The substantially different functional forms of $p(m_s)$ and $p(m_{ns})$ is the consequence of the competition of different cracking mechanisms.

\begin{figure}
 \centering\includegraphics[width=3.4in]{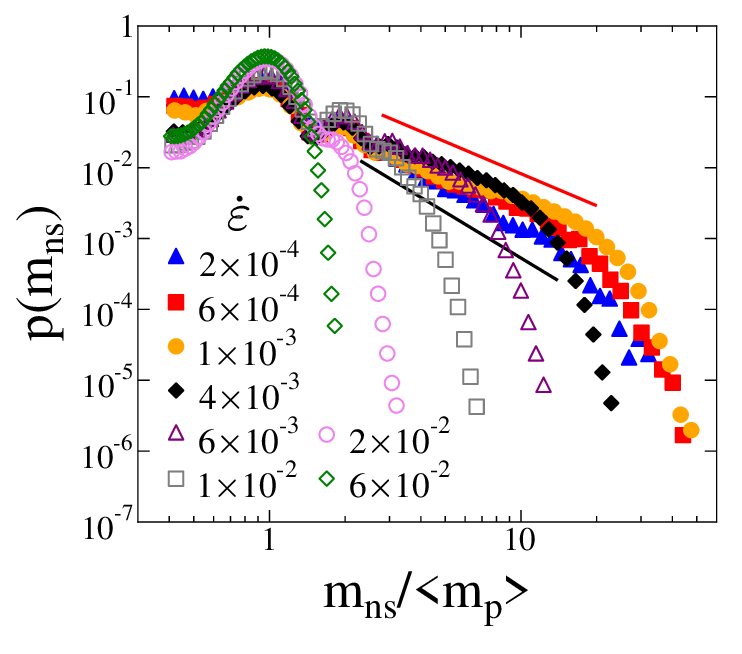}
 \caption{\label{fig:mass_nonspan} Mass distribution of non-spanning fragments $p(m_{ns})$ at the thickness $\Delta R/R=0.05$ for several strain rates $\dot{\varepsilon}$ covering the phases of segmentation, crossover, planar fragmentation, and shattering. $\left<m_p\right>$ denotes the average mass of single polygons. The bold straight lines represent power laws of exponent $\tau =1.5$ and $\tau=2.14$. }
 \end{figure}

\begin{figure}
 \centering\includegraphics[width=3.4in]{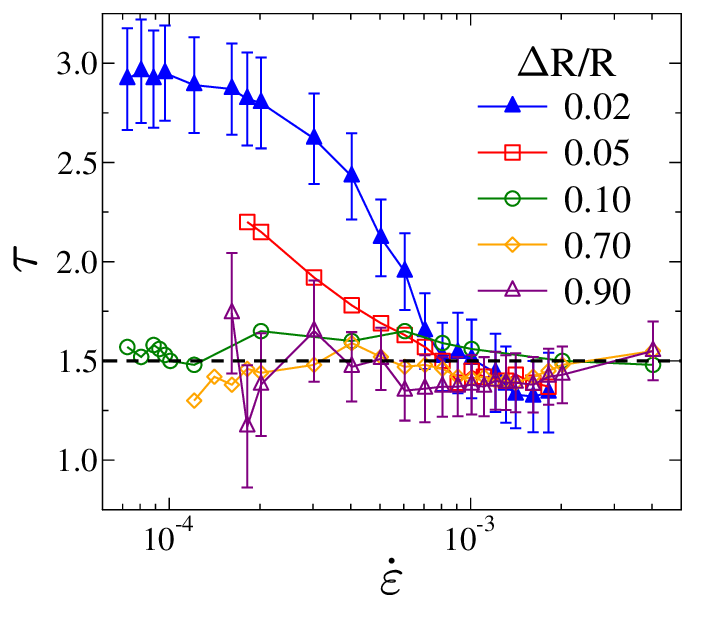}
 \caption{\label{fig:exponents_nonspan} Power law exponent $\tau$ of the mass distribution of non-spanning fragments as a function of the strain rate of explosion $\dot{\varepsilon}$ for several thicknesses. The horizontal dashed line indicates the universal exponent $\tau_{2D}=3/2$ of planar fragmentation. For clarity, the error bars are provided for two thicknesses. }
 \end{figure}

To obtain an overview of the evolution of the statistics of non-spanning fragments, Figure \ref{fig:mass_nonspan} presents their mass distributions $p(m_{ns})$ separately at the ring thickness $\Delta R/R=0.05$ for several strain rates $\dot{\varepsilon}$ covering the phases of segmentation, crossover, planar fragmentation, and shattering of the system. Note that above $\dot{\varepsilon}_f$ no spanning fragments exist so that the distribution $p(m_{ns})$ comprises all the fragments. It can be observed that until spanning fragments exist the slope of the power law regime, i.e.\ the values of the exponent $\tau$ in Eq.\ (\ref{eq:powlaw}) decreases with increasing strain rate $\dot{\varepsilon}$. However, above the fragmentation critical point $\dot{\varepsilon}_f$ the exponent $\tau$ remains practically constant fluctuating around $\tau=3/2$. The behaviour of the power law exponent $\tau$ of non-spanning fragments is summarized in Fig.\ \ref{fig:exponents_nonspan}, where results are presented for several ring thicknesses.
It can be seen that for thin rings such as $\Delta R/R=0.02$ the exponent $\tau$ starts at a relatively high value $\tau\approx 3.0$ at low $\dot{\varepsilon}$, indicating the dominance of small pieces among non-spanning fragments and the rapid decay of the frequency of large ones. \textcolor{black}{However, as larger and larger non-spanning pieces are created by the breakup of the spanning ones with increasing $\dot{\varepsilon}$, the exponent $\tau$ gradually decreases and approaches $\tau_{2D}=3/2$, which is the universal mass distribution exponent of planar fragmentation predicted analytically \cite{astroem_universal_2004,kekalainen_solution_2007} in agreement with experiments \cite{kadono_fragment_1997, oddershede_self-organized_1993, katsuragi_scaling_2003, katsuragi_crossover_2004}}. It is important to emphasize that
once the system enters the phase of planar fragmentation $\tau$ remains practically constant $\tau\approx \tau_{2D}$, apart from numerical error. When the strain rate is increased in the regime $\dot{\varepsilon}>\dot{\varepsilon}_f$, only the cutoff mass $m_0$ of the exponential in Eq.\ (\ref{eq:powlaw}) shifts to lower values. As the ring thickness $\Delta R/R$ increases the deviation of $\tau$ from $\tau_{2D}$ decreases and for sufficiently high thicknesses, where the segmentation phase practically disappears (Fig.\ \ref{fig:phasediag}), the exponent just fluctuates around $\tau_{2D}$ (Fig.\ \ref{fig:exponents_nonspan}).

\section{Transition to shattering}
Further increasing $\dot{\varepsilon}$, a larger and larger fraction of the initial rings gets transformed into powder, and in the limit of very high strain rates, the entire sample gets shattered. The approach towards shattering can be inferred from the mass distributions $p(m_{ns})$ of the few highest strain rates in Fig.\ \ref{fig:mass_nonspan}: the mass range of fragments shrinks in such a way that the cutoff mass $m_0$ of the distributions approaches the average mass of single polygons $\left<m_p\right>$, and eventually a single hump remains at around $m_{ns}/\left<m_p\right>\approx 1$, which is the mass distribution of the powder particles.

\begin{figure}
 \centering\includegraphics[width=3.4in]{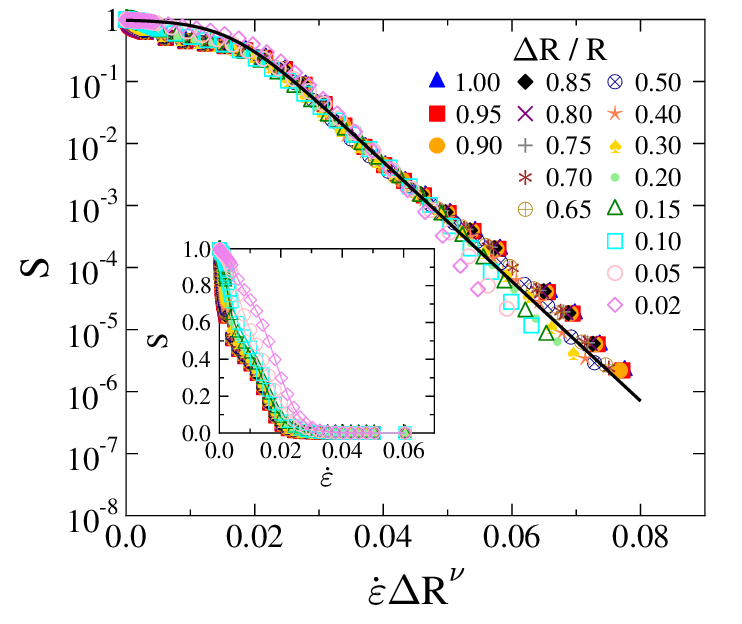}
 \caption{\label{fig:shatter_degree}  $(Inset)$ The mass fraction of non-powder fragments $S$ as a function of strain rate $\dot{\varepsilon}$ for several thicknesses $\Delta R$. $(Main panel)$ Rescaling the data of $(a)$ along the horizontal axis with a proper power $\nu$ of $\Delta R$, curves belonging to different thicknesses collapse on the top of each other. Note the logarithmic scale on the vertical axis. The continuous black line represents a fit with Eq.\ (\ref{eq:scaling_shatt}). }
 \end{figure}

To quantify the approach to the shattering limit, we define the shattering degree $s$ as the mass fraction of powder fragments (single polygons) in the final state \cite{brillantov_physa_2022}, obtained as the total mass of powder pieces divided by the mass of the entire ring. The inset of Figure \ref{fig:shatter_degree} shows that the mass fraction of non-shattered fragments, $S = 1 - s$, decreases from $S \approx 1$ to zero as the strain rate increases for all ring thicknesses $\Delta R/R$. For thicker rings, the curves shift left, indicating faster decay of non-powder fragments. Figure \ref{fig:shatter_degree} demonstrates that rescaling the data of its inset by a power $\nu$ of the ring thickness $\Delta R$ collapses the $S$ curves onto a master curve. The good quality data collapse implies the validity of the scaling form
\begin{align}
S(\dot{\varepsilon}, \Delta R) = \Phi(\dot{\varepsilon}\Delta R^{\nu}),
\label{eq:scaling_s}
\end{align}
where $\Phi(x)$ is the scaling function. \textcolor{black}{Best data collapse was obtained in Fig.\ \ref{fig:shatter_degree} with the exponent $\nu = 0.10 \pm 0.02$.} The semi-log plot reveals that $\Phi(x)$ asymptotically follows an exponential decay described by
\begin{align}
\Phi(x)=A e^{-B(x-x_c)},
\label{eq:scaling_shatt}
\end{align}
where $x_c$ marks the end of the nearly constant regime of $\Phi(x)$ (see Fig.\ \ref{fig:shatter_degree}), and $A, B$ are constants. The scaling structure Eq.\ (\ref{eq:scaling_s}) implies that the characteristic strain rate $\dot{\varepsilon}_s$ of ring shattering has a power law dependence on the ring thickness
\begin{align}
\dot{\varepsilon}_s \sim \Delta R^{-\nu}.
\label{eq:shatter_nu}
\end{align}

\begin{figure}
 \centering\includegraphics[width=3.4in]{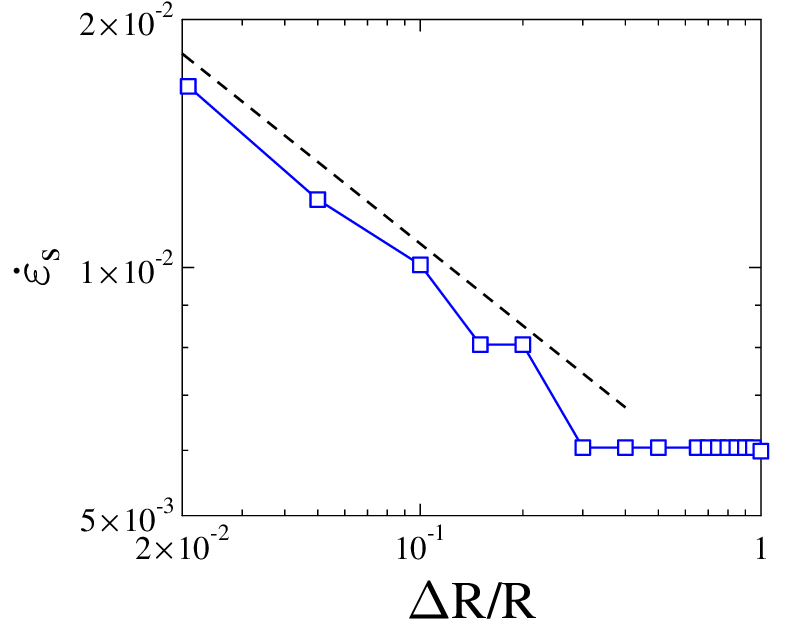}
 \caption{\label{fig:shatter_exp} The shattering threshold $\dot{\varepsilon}_s$ has an approximate power law dependence up to the critical ring thickness $\Delta R_c/R$. The dashed straight line represents a power law Eq.\ (\ref{eq:shatter_nu}) of exponent $\nu=0.33$. }
 \end{figure}
 
To complete the phase diagram of exploding rings (Fig.\ \ref{fig:phasediag}), we determined the shattering critical point $\dot{\varepsilon}_s$ as the strain rate where the mass fraction of powder fragments exceeds that of non-powder pieces. It can be observed in Fig.\ \ref{fig:phasediag} that the phase boundaries of shattering $\dot{\varepsilon}_s$ and planar fragmentation $\dot{\varepsilon}_f$ converge for thin rings $\Delta R/R\to 0$, as expected. \textcolor{black}{The critical strain rate $\dot{\varepsilon}_s$ is also presented separately in Fig.\ \ref{fig:shatter_exp}, which confirms that the scaling law Eq.\ (\ref{eq:shatter_nu}) accurately describes the results with the exponent $\nu=0.33\pm 0.12$ up to the critical ring thickness $\Delta R_c/R$, beyond which $\dot{\varepsilon}_s$ becomes nearly constant. We note that the two values of the exponent $\nu$, derived from the data collapse analysis and from direct fitting, differ beyond their respective uncertainties. This deviation may reflect finite-size effects inherent to the limited extent of the system.}

\section{Discussion}
Breakup processes play a crucial role in various industrial, technological, and scientific applications. In mining and ore processing, controlled fragmentation is essential for efficient size reduction at moderate energy costs \cite{astrom_statistical_2006,bonn_natcomm_2021,turcotte_factals_1986,steacy_automaton_1991,turcotte_fractals_1997,brilliantov_size_2015}. In space exploration, the fragmentation of shell-like rocket components, such as fuel containers, is a major source of space debris, posing significant risks to satellites and manned missions. Similarly, in defense technologies, the ability to control the mass distribution of exploding shells is a key challenge for optimizing performance and safety \cite{space_debris_science_2006,space_debris_impact_2008}.

Fragmentation processes in heterogeneous brittle materials exhibit universal behavior, with power-law mass distributions that remain robust across a wide range of material properties and loading conditions. Experimental and theoretical studies have identified universality classes of fragmentation phenomena, primarily dictated by the dimensionality of the system and the brittle-ductile characteristics of the material \cite{oddershede_self-organized_1993, astrom_universality_2000, astroem_exponential_2004, astroem_universal_2004, wittel_fragmentation_2004, kun_scaling_2006,timar_new_2010}. While these findings allow for the control of global fragment ensemble characteristics, such as average fragment mass, finer details of the fragmentation process require deeper investigation.

In this study, we conducted a comprehensive theoretical analysis of shell fragmentation, using a discrete element model to simulate the explosion of rings in a two-dimensional (2D) embedding space. Our results reveal that the fragmentation behavior of exploding shells is strongly influenced by the structure of the emerging crack pattern at different thicknesses and strain rates. Notably, we identified a dimensional crossover in thin shells, transitioning from one-dimensional (1D) to 2D fragmentation, characterized by distinct scaling laws. At low strain rates, the rings undergo an essentially one-dimensional segmentation, producing spanning fragments that extend between the inner and outer surfaces. As the strain rate increases, internal degrees of freedom become activated, leading to the formation of non-spanning fragments, including surface and bulk pieces.

Our findings demonstrate that spanning fragments follow a Weibull type mass distribution, whereas non-spanning fragments exhibit a power-law mass distribution with an exponential cutoff. Remarkably, the power-law exponent decreases with increasing strain rate until the system enters a fully fragmented state, where no spanning pieces survive. Beyond this fragmentation critical point, a robust power-law mass distribution emerges, with an exponent characteristic of 2D fragmentation phenomena. As the strain rate continues to increase, an increasingly larger fraction of the solid is shattered into powder. Importantly, we identified a threshold strain rate for shattering by observing the exponential decay of non-powder fragments near the transition.

Through extensive simulations, we mapped the phase diagram of the system in the parameter space of strain rate and ring thickness. Our results indicate that while the threshold strain rate for segmentation remains nearly constant, the onset of dimensional crossover and fragmentation shifts to lower strain rates as ring thickness increases. We identified a critical thickness above which the crossover regime between segmentation and fragmentation disappers. Furthermore, we demonstrated that the shattering strain rate follows a power-law decrease with increasing ring thickness.

\textcolor{black}{Comparison of our simulation results with experimental findings is feasible primarily in the limiting cases of thin rings and bulk two-dimensional plates, as no systematic experiments have been conducted to date that vary both ring thickness and strain rate. Nevertheless, it is important to emphasize that our simulations successfully reproduce key qualitative and quantitative features observed in experiments. In particular, we capture the power-law decrease of the average fragment size with increasing strain rate, as well as the functional form of the mass distribution of ring segments, both of which have been reported in laboratory experiments \cite{grady_geometric_1985, ring_warhead_simul_2015} and corroborated by previous simulations \cite{zhou_characteristic_2006, astrom_statistical_2006, ring_warhead_simul_2015}.
For the dynamic fragmentation of brittle plates, extensive experimental studies have shown that the fragment mass distribution follows a power law with an exponential cutoff at large fragment sizes. Remarkably, the power-law exponent is consistently found to be close to 3/2, showing robustness across different materials and loading conditions \cite{kadono_fragment_1997, astrom_exponential_2004, katsuragi_crossover_2004, astrom_statistical_2006}. Our simulations for thick rings reproduce these experimental and theoretical findings well.
Importantly, our results demonstrate that the characteristics observed in the thin-ring limit persist for spanning fragments even in thicker rings. At the same time, the simulations reveal how the fragmentation process evolves as the system transitions between the universality classes of one-dimensional and two-dimensional breakup.}

Based on our 2D analysis, we hypothesize that a similar dimensional crossover arises in three-dimensional (3D) embedding space, where fragmentation transitions between the universality classes of 2D and 3D phenomena. This suggests the possibility of controlling the power-law exponent of the fragment mass distribution by tuning system parameters, which could have practical implications for both industrial and scientific applications.

These findings contribute to the broader understanding of fragmentation dynamics and offer potential strategies for optimizing fragmentation processes across multiple domains. Our results reinforce the significance of dimensionality and material properties in determining fragmentation patterns. Future work should extend this analysis to 3D systems, validating the conjectured crossover behavior and exploring potential engineering applications.

\textcolor{black}{A key aspect of our setup was that the ring thickness $\Delta R$ was varied while keeping the outer radius $R$ fixed. This allowed us to tune the relative thickness $\Delta R / R$ over the range $0 \lesssim \Delta R / R \leq 1$, thereby enabling a controlled crossover from one-dimensional to two-dimensional behavior. To investigate how the size of the sample influences the fracture process at a fixed ring shape, one must fix the ratio of the inner and outer radii, defined as $r = R_{\text{in}} / R$. Varying the outer radius $R$ at constant $r$ ensures geometrical similarity, while the ring area scales as $A = \pi R^2 (1 - r^2)$. This approach allows for finite-size scaling analysis, which can improve the precision of determining, for example, the power-law exponent of the mass distribution of non-spanning fragments. Alternative geometrical setups can also be used to vary the ring thickness, such as increasing the outer radius while keeping the inner radius $R_{\text{in}}$ fixed. However, since the ratio $r$ uniquely determines the ring's shape, such cases can be mapped onto our setup through geometrical similarity and are thus expected to yield equivalent results.}

\textcolor{black}{A dimensional crossover from one- to two-dimensional fragmentation behavior can also be achieved by increasing the ring's extension perpendicular to its plane. In this way, a thin ring gradually becomes a cylindrical shell, which can be fragmented by initiating an explosion along its axis. This scenario has been investigated experimentally through the explosive fragmentation of thin ceramic tubes, where the imparted energy was systematically varied \cite{katsuragi_explosive_2005}. The experiments showed that at sufficiently high energies, the mass distribution of fragments exhibits a power-law decay followed by an exponential cutoff. However, as this setup represents a fully three-dimensional fragmentation process, it lies outside the scope of our current model.}

\textcolor{black}{In the present study we focused on the evolution of the statistics of different types of fragments, to understand the 1D to 2D transition occurring in the breakup of rings. Fragments are defined as material pieces that are fully enclosed by cracks and the free surface of the original sample. Consequently, the transition from segmentation to fragmentation, and ultimately to shattering, is marked by significant changes in the structure of the underlying crack pattern. To develop a comprehensive understanding of this evolution, we have to further investigate how the geometrical features of individual micro-cracks and extended macro-cracks—composed of multiple micro-cracks—vary with increasing strain rate across samples of different thicknesses. The evolution and structure of macro-cracks may exhibit interesting analogies to invasion percolation \cite{wilkinson_invasion_1983} which will be explored in a forthcoming publications.}












\bibliographystyle{cas-model2-names}

\bibliography{statphys_fracture}



\end{document}